\documentclass[final,5p,times,twocolumn]{elsarticle}

\usepackage{graphicx}
\usepackage{dcolumn}
\usepackage{bm}
\usepackage{amsfonts}
\usepackage{latexsym}
\usepackage{amssymb}
\usepackage{verbatim}
\usepackage{amsthm}
\usepackage{amsmath}
\usepackage{mathrsfs}
\usepackage{wrapft}

\newcommand{\SF}{{{\mathbb S}}}

\journal{Physics Letter A}

\begin{document}

\begin{frontmatter}



\title{
  Vacuum fluctuations and balanced homodyne detection \\
  through ideal multi-mode photon number or power counting detectors
}


\author{Kouji Nakamura}
\ead{kouji.nakamura@nao.ac.jp}


\address{
  Gravitational-Wave Science Project,
  National Astronomical Observatory,
  Mitaka, Tokyo 181-8588, Japan
}

\begin{abstract}
  The balanced homodyne detection as a readout scheme of
  gravitational-wave detectors is carefully examined, which specifies
  the directly measured quantum operator in the detection.
  This specification is necessary to apply the quantum measurement
  theory to gravitational-wave detections.
  We clarify the contribution of vacuum fluctuations to the noise
  spectral density without using the two-photon formulation.
  We found that the noise spectral density in the two-photon
  formulation includes vacuum fluctuations from the main
  interferometer but does not includes those from the local
  oscillator which depends on the directly measured operators.
\end{abstract}

\begin{keyword}
quantum measurement theory \sep
vacuum fluctuations \sep
balanced homodyne detection \sep
gravitational-wave detectors




\end{keyword}

\end{frontmatter}




\section{Introduction}
\label{sec:Introduction}


One of the motivations of the recent quantum measurement
theory~\cite{Ozawa-2004} is gravitational-wave detection.
However, the actual application of this theory to the
gravitational-wave detection requires its extension to the quantum
field theory.
Furthermore, in quantum measurement theory, we have to specify the
directly measured quantum operator.
In interferometric gravitational-wave detectors, we may regard  that
the directly measured operator is specified at their ``{\it readout
  scheme}'' in the detectors.
The readout scheme in gravitational-wave detectors is the optical
system to specify the optical fields which are detected at the
photodetectors through the signal output from the main-interferometer and
the reference field which injected from the ``{\it local oscillator}.''
The research on this readout scheme is important for the development
and application of the quantum measurement theory to
gravitational-wave detections.


Current gravitational-wave detectors use the DC readout scheme, in
which the output photon power is directly measured.
On the other hand, ``{\it homodyne detections}'' are regarded as one
of candidates of the readout scheme in the near
future~\cite{S.Steinlechner-et-al-2015-T.Zhang-2017}.
In
Ref.~\cite{H.J.Kimble-Y.Levin-A.B.Matsko-K.S.Thorne-S.P.Vyatchanin-2001},
it is written that the output quadrature $\hat{b}_{\theta}$ defined by
\begin{eqnarray}
  \label{eq:DBHD_20180805_1.1}
  \hat{b}_{\theta}
  :=
  \cos\theta \hat{b}_{1} + \sin\theta \hat{b}_{2}
\end{eqnarray}
is measured by the ``{\it balanced homodyne detection}.''
The output operator $\hat{b}_{\theta}$ includes gravitational-wave
signal $h(\Omega)$ as
\begin{eqnarray}
  \label{eq:DBHD_20180805_1.2}
  \hat{b}_{\theta}
  &=&
      R(\Omega,\theta)\left(
      \hat{h}_{n}(\Omega,\theta) + h(\Omega)
      \right),
\end{eqnarray}
where $\theta$ is the homodyne angle, $\hat{h}_{n}(\Omega)$ is the
noise operator which is given by the linear combination of the
annihilation and creation operators of photons injected to the main
interferometer.
However, it does not seem that there is clear description on the
actual measurement processes of the operator
(\ref{eq:DBHD_20180805_1.1}) or directly measured quantum operators in
these processes.


For this motivation, in
Refs.~\cite{K.Nakamura-M.-K.Fujimoto-double-balanced},
we examined the case where the directly measured operators are the
number operator:
\begin{eqnarray}
  \label{eq:each-mode-photon-number-def}
  \hat{n}(\omega) = \hat{a}^{\dagger}(\omega)\hat{a}(\omega)
\end{eqnarray}
using the annihilation [$\hat{a}(\omega)$] and creation
[$\hat{a}^{\dagger}(\omega)$] operators of the electric field.
From the usual commutation relation
\begin{eqnarray}
  &&
     \label{eq:usual-commutation-relation-0}
     \left[\hat{a}(\omega),\hat{a}(\omega')\right]
     =
     \left[\hat{a}^{\dagger}(\omega),\hat{a}^{\dagger}(\omega')\right]
     =
     0
     ,
     \\
  &&
     \left[\hat{a}(\omega),\hat{a}^{\dagger}(\omega')\right]=2\pi\delta(\omega-\omega')
     ,
     \label{eq:usual-commutation-relation-vacuum}
\end{eqnarray}
the eigenvalue of the operator $\hat{n}(\omega)$ becomes countable
number.
This countable number give rise to the notion of ``photon'' and we can
count this number.
As a result of this examination, we reached to a conclusion that we
cannot measure the expectation value of the operator
(\ref{eq:DBHD_20180805_1.1}) by the balanced homodyne
detection~\cite{E.Shchukin-Th.Richter-W.Vogel-2005}.
However, the operator (\ref{eq:each-mode-photon-number-def}) is not
appropriate as a directly measured operator in gravitational-wave
detectors, since multi-mode detections are essential in these detectors.


As a directly measured operators in multi-mode detections, Glauber's
photon number operator
\begin{eqnarray}
  \label{eq:multi-mode-photon-number-def}
  \hat{N}_{a}(t)
  :=
  \frac{\kappa_{n}c}{2\pi\hbar} {\cal A}
  \hat{E}_{a}^{(-)}(t)\hat{E}_{a}^{(+)}(t).
\end{eqnarray}
is often used in many literatures.
The factor $\frac{\kappa_{n}c}{2\pi\hbar} {\cal A}$ in
Eq.~(\ref{eq:multi-mode-photon-number-def}) is chosen for our
convention and the coefficients $\kappa_{n}$ is a phenomenological
parameter whose dimension is [time] which includes ``quantum
efficiency''.
This factor is not important within our discussions.
On the other hand, the output electric field $\hat{E}_{a}$ are
separated into their positive- and negative-frequency part as
\begin{eqnarray}
  \label{eq:hatEa-positive-negative-separation}
  \hat{E}_{a}(t)&=&\hat{E}_{a}^{(+)}(t)+\hat{E}_{a}^{(-)}(t), \quad
  \hat{E}_{a}^{(-)}(t)=\left[\hat{E}_{a}^{(+)}(t)\right]^{\dagger}, \\
  \label{eq:electric-field-notation-total-electric-field-positive}
  \hat{E}_{a}^{(+)}(t)
  &=&
  \int_{0}^{+\infty} \frac{d\omega}{2\pi}
  \sqrt{\frac{2\pi\hbar\omega}{{\cal A}c}}
  \hat{a}(\omega) e^{-i\omega t}
  .
\end{eqnarray}
Here, ${\cal A}$ in Eqs.~(\ref{eq:multi-mode-photon-number-def}) and
(\ref{eq:electric-field-notation-total-electric-field-positive}) is
the sectional area of the laser beam.
The operator (\ref{eq:multi-mode-photon-number-def}) is an extension
of the operator (\ref{eq:each-mode-photon-number-def}) to multi-mode
cases within the Maxwell theory~\cite{R.J.Glauber-1963-130}.


In quantum measurement theories of optical fields, there was a long
controversy on which variable is the directly measured by
photodetectors in multi-mode
cases~\cite{R.J.Glauber-1963-130,multi-mode-history,H.J.Kimble-L.Mandel-1984}.
Some insisted that the directly measured operators at photodetectors
is the above Glauber's photon number operator
(\ref{eq:multi-mode-photon-number-def}), and some insisted that the
direct observable of photodetectors is the power of the optical field.
Within this history, Kimble and Mandel~\cite{H.J.Kimble-L.Mandel-1984}
pointed out that we cannot say neither, in general.
The current consensus of this issue will be that the directly measured
operator at the photodetectors depends on the physical properties of
the photodetectors such as their band structure for photo-electron
emission, in general.
In the gravitational-wave community, it is often said that the
probability of the excitation of the photocurrent is proportional to
the power operator
\begin{eqnarray}
  \label{eq:multi-mode-power-def}
  \hat{P}_{a}(t)
  :=
  \frac{\kappa_{p}c}{4\pi\hbar} {\cal A}
  \left(\hat{E}_{a}(t)\right)^{2}
\end{eqnarray}
of the measured optical field.
The factor $\frac{\kappa_{p}c}{4\pi\hbar} {\cal A}$ in
Eq.~(\ref{eq:multi-mode-power-def}) is chosen for our convention and
the coefficient $\kappa_{p}$ is a phenomenological parameter whose
dimension is [time] which includes ``quantum efficiency''.
This is not important within our discussions as in the case of
Eq.~(\ref{eq:multi-mode-photon-number-def}).
It is also true that there are many literatures in which the
photo-detection is treated as a classical stochastic process, in which
the detection probability is proportional to the expectation value of
the power
operator~\cite{Power-counting-explanation}.


In this Letter, we examine these two cases where the directly measured
operators at the photodetector is Glauber's number operator
(\ref{eq:multi-mode-photon-number-def}) and that is the power operator
(\ref{eq:multi-mode-power-def}).
Furthermore, we estimate the quantum noise in these two cases.
In many literatures of the gravitational-wave detection, it is written
that the single sideband noise spectral density
$\bar{S}_{A}^{(s)}(\omega)$ for an arbitrary operator
$\hat{A}(\omega)$ with the vanishing expectation value in the
``{\it stationary}'' system is given by
\begin{eqnarray}
  \label{eq:Kimble-noise-spectral-density-single-side}
  \!\!\!\!\!\!\!\!
  \frac{1}{2}
  2\pi \delta(\omega-\omega') \bar{S}_{A}^{(s)}(\omega)
  :=
  \frac{1}{2}
  \langle
  \hat{A}(\omega)\hat{A}^{\dagger}(\omega')
  +
  \hat{A}^{\dagger}(\omega')\hat{A}(\omega)
  \rangle
  .
\end{eqnarray}
This noise spectral density is introduced by Kimble et al. in Ref.~\cite{H.J.Kimble-Y.Levin-A.B.Matsko-K.S.Thorne-S.P.Vyatchanin-2001}
in the context of the two-photon
formulation~\cite{C.M.Caves-B.L.Schumaker-1985}.
This two-photon formulation is commonly used in the community of
gravitational-wave detection.
However, in this Letter, we do not use the two-photon formulation,
though some final formulae are written in terms of the two-photon
formulation.
We also examine the original meaning of Kimble's noise spectral
density (\ref{eq:Kimble-noise-spectral-density-single-side}) and
derive the deviation from this noise formula
(\ref{eq:Kimble-noise-spectral-density-single-side}).


\section{Balanced Homodyne Detections by multi-mode detectors}
\label{sec:Homodyne_Detections_by_multi-mode_detectors}


Here, we examine the expectation value of the balanced homodyne detection.
In Sec.~\ref{sec:Homodyne_Detections_by_photon-number-counting}, this
expectation value is evaluated under the premise that the directly
measured operator at the photodetector is Glauber's number operator
(\ref{eq:multi-mode-photon-number-def}).
In Sec.~\ref{sec:Homodyne_Detections_by_power-counting}, we discuss
the expectation value under the premise that the directly measured
operator is the power operator (\ref{eq:multi-mode-power-def}).


\subsection{Balanced Homodyne Detections by Glauber's Photon-Number Counting Detectors}
\label{sec:Homodyne_Detections_by_photon-number-counting}

\begin{figure}[ht]
  \centering
  \includegraphics[width=0.48\textwidth]{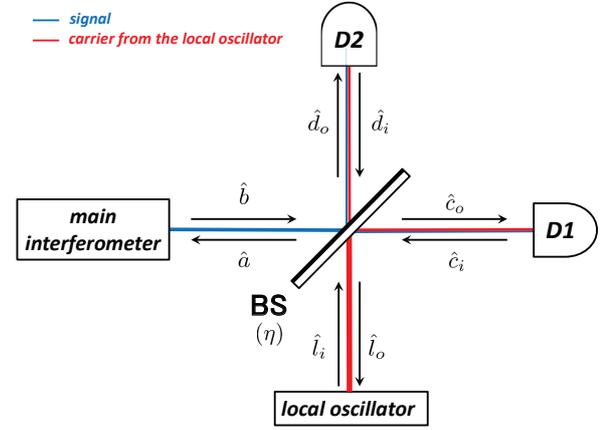}
  \caption{
    Configuration of the interferometer for the balanced homodyne
    detection. Notations of the quadrature for the electric fields in
    the main text is also described. 
  }
  \label{fig:Balanced_homodyne_detection}
\end{figure}

Here, we review the balanced homodyne detection depicted in
Fig.~\ref{fig:Balanced_homodyne_detection}.
Throughout this Letter, we want to measure the signal field
$\hat{E}_{b}(t)$.

The electric field from the local oscillator $\hat{E}_{l_{i}}$ is in
the coherent state $|\gamma\rangle$ with
$\hat{l}_{i}(\omega)|\gamma\rangle_{l_{i}}=\gamma(\omega)|\gamma\rangle_{l_{i}}$.
In the time domain, the state $|\gamma\rangle$ satisfies
\begin{eqnarray}
  \label{eq:coherent-state-def-li-time-domain}
  \hat{E}_{l_{i}}^{(+)}(t)|\gamma\rangle_{l_{i}}
  &=&
  \sqrt{\frac{2\pi\hbar}{{\cal A}c}} \gamma(t) |\gamma\rangle_{l_{i}}
  ,
  \\
  \label{eq:gammat-gammaomega-relation}
  \gamma(t)
  &:=&
  \int_{0}^{\infty} \frac{d\omega}{2\pi}
  \sqrt{|\omega|}
  \gamma(\omega)
  e^{-i\omega t}
  .
\end{eqnarray}
The output signal field $\hat{E}_{b}(t)$ and the field
$\hat{E}_{l_{i}}(t)$ from the local oscillator is mixed through the beam splitter
with the transmissivity $1/2$.


At the beam splitter in Fig.~\ref{fig:Balanced_homodyne_detection},
the fields $\hat{E}_{b}(t)$ and $\hat{E}_{l_{i}}(t)$ are transformed
to the fields $\hat{E}_{c_{o}}(t)$ and $\hat{E}_{d_{o}}(t)$ as
\begin{eqnarray}
  \label{eq:DBHD_20180805_2.10}
  \hat{E}_{c_{o}}(t)
  &=&
      \frac{1}{\sqrt{2}} \hat{E}_{b}(t) + \frac{1}{\sqrt{2}} \hat{E}_{l_{i}}(t)
      ,
  \\
  \label{eq:hatEdo-hatEb-hatEli-beamsplitter-relation-BHD}
  \hat{E}_{d_{o}}(t)
  &=&
  - \frac{1}{\sqrt{2}} \hat{E}_{l_{i}}(t)
  +\frac{1}{\sqrt{2}} \hat{E}_{b}(t)
  .
\end{eqnarray}


The electric fields $\hat{E}_{d_{i}}$ and $\hat{E}_{c_{i}}(t)$ are in
their vacua
\begin{eqnarray}
  \label{eq:di-ci-vacuum-def}
  \hat{d}_{i}(\omega)|0\rangle_{d_{i}}
  =
  \hat{c}_{i}(\omega)|0\rangle_{c_{i}}
  =
  0
  ,
\end{eqnarray}
respectively, and the field $\hat{E}_{a}(t)$ are given by
$\hat{E}_{d_{i}}$ and $\hat{E}_{c_{i}}(t)$ through the beam splitter
condition.
In general, the state of the field $\hat{E}_{b}$ depends on the state
of the input field $\hat{E}_{a}$ and the other optical fields which is
injected to the main
interferometer~\cite{H.J.Kimble-Y.Levin-A.B.Matsko-K.S.Thorne-S.P.Vyatchanin-2001}.
Furthermore, we consider the situation where the output electric field
$\hat{E}_{b}$ includes the information of classical forces as in
Eq.~(\ref{eq:DBHD_20180805_1.2}) and this information are
measured through the expectation value of the operator $\hat{E}_{b}$.
To evaluate the expectation value of the signal field $\hat{E}_{b}$,
we have to specify the state of the total system as
\begin{eqnarray}
  \label{eq:DBHD_20180805_2.14}
  |\Psi\rangle
  =
  |\gamma\rangle_{i_{i}}\otimes|0\rangle_{c_{i}}\otimes|0\rangle_{d_{i}}\otimes|\psi\rangle_{main}.
\end{eqnarray}
Here, the state $|\psi\rangle_{main}$ is for the electric fields
associated with the main interferometer, which is independent
of the state $|\gamma\rangle_{l_{i}}$, $|0\rangle_{c_{i}}$, and
$|0\rangle_{d_{i}}$.
The expectation value of the field $\hat{E}_{a}$ means
\begin{eqnarray}
  \label{eq:DBHD_20180805_2.16}
  \langle\hat{E}_{a}\rangle
  :=
  \langle\Psi|\hat{E}_{a}|\Psi\rangle.
\end{eqnarray}


Here, we regard that Glauber's photon number operators
\begin{eqnarray}
  \label{eq:balanced-homodyne-direct-observable-number-D1}
  \hat{N}_{c_{o}}(t)
  &:=&
       \frac{\kappa_{n}c}{2\pi\hbar} {\cal A}
       \hat{E}_{c_{o}}^{(-)}(t)\hat{E}_{c_{o}}^{(+)}(t)
       ,
  \\
  \label{eq:balanced-homodyne-direct-observable-number-D2}
  \hat{N}_{d_{o}}(t)
  &:=&
       \frac{\kappa_{n}c}{2\pi\hbar} {\cal A}
       \hat{E}_{d_{o}}^{(-)}(t)\hat{E}_{d_{o}}^{(+)}(t)
\end{eqnarray}
are directly measured at the photodetector D1 and D2 in
Fig.~\ref{fig:Balanced_homodyne_detection}, respectively.
Substituting Eq.~(\ref{eq:DBHD_20180805_2.10}) into
Eq.~(\ref{eq:balanced-homodyne-direct-observable-number-D1}), we
obtain
\begin{eqnarray}
  \hat{N}_{c_{o}}(t)
  \!\!\!\!&=&\!\!\!\!
      \frac{1}{2} \hat{N}_{b}(t)
      +
      \frac{1}{2} \hat{N}_{l_{i}}(t)
      \nonumber\\
  &&
     +
     \frac{1}{2} \kappa_{n} \frac{{\cal A}c}{2\pi\hbar}
     \left(
     \hat{E}_{l_{i}}^{(-)}(t) \hat{E}_{b}^{(+)}(t)
     +
     \hat{E}_{b}^{(-)}(t) \hat{E}_{l_{i}}^{(+)}(t)
     \right)
     ,
     \label{eq:simple-homodyne-direct-observable-number-b-li}
\end{eqnarray}
while the substitution of 
Eq.~(\ref{eq:hatEdo-hatEb-hatEli-beamsplitter-relation-BHD}) into
Eq.~(\ref{eq:balanced-homodyne-direct-observable-number-D2}) yields
\begin{eqnarray}
  \hat{N}_{d_{o}}(t)
  \!\!\!\!&=&\!\!\!\!
      \frac{1}{2} \hat{N}_{b}(t)
      +
      \frac{1}{2} \hat{N}_{l_{i}}(t)
      \nonumber\\
  &&
     -
     \frac{1}{2} \kappa_{n} \frac{{\cal A}c}{2\pi\hbar}
     \left(
     \hat{E}_{l_{i}}^{(-)}(t) \hat{E}_{b}^{(+)}(t)
     +
     \hat{E}_{b}^{(-)}(t) \hat{E}_{l_{i}}^{(+)}(t)
     \right)
     .
     \label{eq:D2-balanced-homodyne-direct-observable-number-b-li}
\end{eqnarray}


From the expectation values of
Eqs.~(\ref{eq:simple-homodyne-direct-observable-number-b-li}) and (\ref{eq:D2-balanced-homodyne-direct-observable-number-b-li}), we
obtain
\begin{eqnarray}
  \label{eq:balanced-homodyne-direct-observable-number-balanced-exp}
  \!\!\!\!
  \frac{1}{\kappa_{n}}
  \left\langle
  \hat{N}_{c_{o}}(t)
  -
  \hat{N}_{d_{o}}(t)
  \right\rangle
  =
  \frac{{\cal A}c}{2\pi\hbar}
  \left\langle
  \gamma^{*}(t) \hat{E}_{b}^{(+)}(t)
  +
  \gamma(t) \hat{E}_{b}^{(-)}(t)
  \right\rangle
  ,
\end{eqnarray}
and we define the signal operator $\hat{s}_{N}(t)$ as
\begin{eqnarray}
  \hat{s}_{N}(t)
  &:=&
       \frac{1}{\kappa_{n}}
       \left[
       \hat{N}_{c_{o}}(t)
       - \hat{N}_{d_{o}}(t)
       \right]
       \label{eq:BHD-signal-operator-number-time-domain-def}
  \\
  &=&
      \frac{{\cal A}c}{2\pi\hbar} \left[
      \hat{E}_{l_{i}}^{(-)}(t) \hat{E}_{b}^{(+)}(t)
      +
      \hat{E}_{b}^{(-)}(t) \hat{E}_{l_{i}}^{(+)}(t)
      \right]
      \label{eq:hatsNt-electric-field}
 \end{eqnarray}
so that
\begin{eqnarray}
  \langle \hat{s}_{N}(t) \rangle
  &=&
      \sqrt{\frac{{\cal A}c}{2\pi\hbar}}
      \left[
      \gamma^{*}(t)
      \left\langle \hat{E}_{b}^{(+)}(t) \right\rangle
      +
      \gamma(t)
      \left\langle \hat{E}_{b}^{(-)}(t) \right\rangle
      \right]
      .
      \label{eq:sNt-exp-valu}
\end{eqnarray}
We note that $\hat{s}_{N}(t)$ is a self-adjoint operator.


Here, we consider the monochromatic local oscillator case, in which
the complex amplitude $\gamma(\omega)$ in
Eq.~(\ref{eq:gammat-gammaomega-relation}) is given by
\begin{eqnarray}
  \label{eq:monochromatic-gamma-omega}
  \gamma(\omega) = 2\pi\gamma\delta(\omega-\omega_{0}),
  \quad
  \omega_{0}>0,
  \quad
  \gamma=:|\gamma|e^{+i\theta}
\end{eqnarray}
and consider the situation $\omega_{0}\gg\omega>0$.
In this case, the Fourier transformation of the expectation value
(\ref{eq:sNt-exp-valu}) is given by
\begin{eqnarray}
  \langle\hat{s}_{N}(\omega)\rangle
  \sim
  \omega_{0}
  |\gamma|
  \left\langle
  e^{-i\theta}
  \hat{b}(\omega_{0}+\omega)
  +
  e^{+i\theta}
  \hat{b}^{\dagger}(\omega_{0}-\omega)
  \right\rangle
  .
  \label{eq:middle-simple-homodyne-direct-observable-number-b-li-exp-tmp2}
\end{eqnarray}
We note that $\omega_{0}$ in
Eq.~(\ref{eq:middle-simple-homodyne-direct-observable-number-b-li-exp-tmp2})
is just the central frequency of the local oscillator and have nothing
to do with the central frequency of the signal field $\hat{E}_{b}(t)$.
Therefore,
Eq.~(\ref{eq:middle-simple-homodyne-direct-observable-number-b-li-exp-tmp2})
is still valid even in the case ``{\it heterodyne detection}.''


Now, we choose $\omega_{0}$ so that this frequency coincides with the
central frequency of the signal field $\hat{E}_{b}(t)$.
This is the ``{\it homodyne detection}''.
Then, we may identify the quadratures $\hat{b}(\omega_{0}+\omega)$ and
$\hat{b}(\omega_{0}-\omega)$ with the upper- and lower-sideband
quadratures $\hat{b}_{+}(\omega)$ and $\hat{b}_{-}(\omega)$ in the
two-photon formulation~\cite{C.M.Caves-B.L.Schumaker-1985},
respectively.
We may introduce the amplitude quadrature $\hat{b}_{1}(\omega)$ and
the phase quadrature $\hat{b}_{2}(\omega)$ by
\begin{eqnarray}
  \label{eq:hatb1-hatb2-def}
  \hat{b}_{1}
  :=
  \frac{1}{\sqrt{2}}\left(\hat{b}_{+}+\hat{b}_{-}^{\dagger}\right)
  , \quad
  \hat{b}_{2}
  :=
  \frac{1}{\sqrt{2}i}\left(\hat{b}_{+}-\hat{b}_{-}^{\dagger}\right)
  .
\end{eqnarray}
In terms of these quadratures $\hat{b}_{1,2}(\omega)$
Eq.~(\ref{eq:middle-simple-homodyne-direct-observable-number-b-li-exp-tmp2})
is given by
\begin{eqnarray}
  \frac{1}{\sqrt{2}\omega_{0}|\gamma|}
  \langle\hat{s}_{N}(\omega)\rangle
  \sim
  \left\langle
  \hat{b}_{\theta}(\omega)
  \right\rangle
  .
  \label{eq:hatsnomega-expvalue-delta-gamma-sideband-picture-result}
\end{eqnarray}
Thus, when Glauber's photon number operator is the directly measured
operator at the photodetectors, we can measure expectation value of
the operator $\hat{b}_{\theta}$.


\subsection{Balanced Homodyne Detections by Power Counting Detectors}
\label{sec:Homodyne_Detections_by_power-counting}


Through the conditions (\ref{eq:DBHD_20180805_2.10}) and
(\ref{eq:hatEdo-hatEb-hatEli-beamsplitter-relation-BHD}) at the beam
splitter and the definition (\ref{eq:multi-mode-power-def}) of the
operator $\tilde{P}_{a}(t)$, the power operators $\hat{P}_{c_{o}}(t)$
at the D1 and $\hat{P}_{d_{o}}(t)$ at D2 are given by
\begin{eqnarray}
  \hat{P}_{c_{o}}(t)
  &=&
      \frac{1}{2} \hat{P}_{b}(t) + \frac{1}{2} P_{l_{i}}(t)
      \nonumber\\
  &&
      +
      \frac{1}{2} \kappa_{p} \frac{{\cal A}c}{2\pi\hbar} \left(
      \hat{E}_{b}(t) \hat{E}_{l_{i}}(t) + \hat{E}_{l_{i}}(t) \hat{E}_{b}(t)
      \right)
     ,
     \label{eq:balanced-homodyne-direct-observable-power-D1}
  \\
  \hat{P}_{d_{o}}(t)
  &=&
      \frac{1}{2} \hat{P}_{b}(t) + \frac{1}{2} \hat{P}_{l_{i}}(t)
      \nonumber\\
  &&
      -
      \frac{1}{2} \kappa_{p} \frac{{\cal A}c}{2\pi\hbar} \left(
      \hat{E}_{b}(t) \hat{E}_{l_{i}}(t) + \hat{E}_{l_{i}}(t) \hat{E}_{b}(t)
      \right)
      .
     \label{eq:balanced-homodyne-direct-observable-power-D2}
\end{eqnarray}
As Sec.~\ref{sec:Homodyne_Detections_by_photon-number-counting}, we
define the signal operator $\hat{s}_{P}(t)$ by
\begin{eqnarray}
  \hat{s}_{P}(t)
  &:=&
       \frac{1}{2\kappa_{p}} \left[
       \hat{P}_{c_{o}}(t)
       -
       \hat{P}_{d_{o}}(t)
       \right]
       \label{eq:hatsPt-def}
  \\
  &=&
      \frac{{\cal A}c}{4\pi\hbar} \left[
      \hat{E}_{l_{i}}(t) \hat{E}_{b}(t) + \hat{E}_{b}(t) \hat{E}_{l_{i}}(t)
      \right]
      \label{eq:hatsPt-field-expression}
      .
\end{eqnarray}


To evaluate the expectation value of the signal operator
$\hat{s}_{P}(t)$, we assume the commutation relation
\begin{eqnarray}
  \label{eq:hatEb-hatEli-commutation-relation}
  \left[\hat{E}_{b}(t),\hat{E}_{l_{i}}(t)\right]=0,
\end{eqnarray}
which is justified in
Ref.~\cite{K.Nakamura-BHD-Multi-Full-Paper-in-preparation}.
Then, we obtain
\begin{eqnarray}
  \label{eq:hatsPt-exp-valu}
  \left\langle\hat{s}_{P}(t)\right\rangle
  =
  \sqrt{\frac{{\cal A}c}{2\pi\hbar}}
  \left(\gamma(t)+\gamma^{*}(t)\right)
  \left\langle\hat{E}_{b}(t)\right\rangle
  .
\end{eqnarray}
In the case of the monochromatic local oscillator
(\ref{eq:monochromatic-gamma-omega}) and the situation
$\omega_{0}\gg\omega>0$, the Fourier transformation of
Eq.~(\ref{eq:hatsPt-exp-valu}) is
\begin{eqnarray}
  \left\langle\hat{s}_{P}(\omega)\right\rangle
  :=
  \int_{-\infty}^{+\infty} dt e^{+i\omega t}
  \left\langle\hat{s}_{P}(t)\right\rangle
  \sim
  \sqrt{2} \omega_{0} |\gamma|
  \left\langle
  \hat{b}_{\theta}(\omega)
  \right\rangle
  .
  \label{eq:hatsPt-exp-valu-Fourier-monochro-local-omega0gtromegagrt0}
\end{eqnarray}
This is the same result as
Eq.~(\ref{eq:hatsnomega-expvalue-delta-gamma-sideband-picture-result}).


\section{Noise Spectral Densities}
\label{sec:Noise-spectral-densities}


In the two-photon formulation, sideband fluctuations in the frequency
$\omega_{0}\pm\omega$ with the central frequency $\omega_{0}$ is
considered and Kimble's single-sideband noise-spectral
density (\ref{eq:Kimble-noise-spectral-density-single-side}) is
commonly used.
The ``{\it single sideband}'' means the evaluation of noises only in
the frequency range $\omega>0$ of the positive- and negative-sideband
$\omega_{0}\pm\omega$.
The frequencies $\omega$ and $\omega'$ in
Eq.~(\ref{eq:Kimble-noise-spectral-density-single-side}) is the
sideband frequencies in the two-photon formulation.
If we consider the noise in both sideband $\omega\gtrless 0$, the
noise spectral density $\bar{S}_{A}^{(d)}(\omega)$ is call
``{\it double sideband}''
\begin{eqnarray}
  \label{eq:Kimble-noise-spectral-density-double-side}
  \!\!\!\!\!
  2\pi \delta(\omega-\omega') \bar{S}_{A}^{(d)}(\omega)
  :=
  \frac{1}{2} \langle
  \hat{A}(\omega)\hat{A}^{\dagger}(\omega')
  +
  \hat{A}^{\dagger}(\omega')\hat{A}(\omega)
  \rangle
  .
\end{eqnarray}
Furthermore, (double sideband) ``{\it correlation spectral density}''
of operators $\hat{A}(\omega)$ and $\hat{B}(\omega)$ with
$\langle\hat{A}(\omega)\rangle=\langle\hat{B}(\omega)\rangle=0$ is
\begin{eqnarray}
  \label{eq:Kimble-noise-correlation-spectral-density-double-side}
  \!\!\!\!\!
  2\pi \delta(\omega-\omega') \bar{S}_{AB}^{(d)}(\omega)
  :=
  \frac{1}{2} \langle
  \hat{A}(\omega)\hat{B}^{\dagger}(\omega')
  +
  \hat{B}^{\dagger}(\omega')\hat{A}(\omega)
  \rangle
  .
\end{eqnarray}


To examine the meaning of the correlation
(\ref{eq:Kimble-noise-correlation-spectral-density-double-side}), we
consider the time-domain expression of this formulae for the
correlation spectral density through the Fourier transformation.
Introducing the time-domain variables $\hat{A}(t)$ and $\hat{B}(t)$ as
\begin{eqnarray}
  \label{eq:time-domain-A}
  \hat{A}(t)
  \!\!\!\!&=&\!\!\!\!
      \int_{-\infty}^{+\infty} \frac{d\omega}{2\pi}
      \left(
      \Theta(\omega) \hat{A}(\omega)
      +
      \Theta(-\omega) \hat{A}^{\dagger}(-\omega)
      \right)
      e^{-i\omega t}
      ,
  \\
  \label{eq:time-domain-B}
  \hat{B}(t)
  \!\!\!\!&=&\!\!\!\!
      \int_{-\infty}^{+\infty} \frac{d\omega}{2\pi}
      \left(
      \Theta(\omega) \hat{B}(\omega)
      +
      \Theta(-\omega) \hat{B}^{\dagger}(-\omega)
      \right)
      e^{-i\omega t}
      ,
\end{eqnarray}
Eq.~(\ref{eq:Kimble-noise-correlation-spectral-density-double-side})
yields
\begin{eqnarray}
  \label{eq:Kimble-noise-correlation-function-time-domain-tau}
  \bar{C}_{AB}(\tau)
  =
  \frac{1}{2} \left\langle
  \hat{A}(t+\tau)\hat{B}(t)
  +
  \hat{B}(t)\hat{A}(t+\tau)
  \right\rangle
  .
\end{eqnarray}
Note that the left-hand side of
Eq.~(\ref{eq:Kimble-noise-correlation-function-time-domain-tau})
depends only on ``$\tau$'', while the right-hand side may depend both
on ``$t$'' and ``$\tau$'', in general.
This dependence implies the ``{\it stationarity}'' of the correlation.
If we consider the non-stationary cases, the correlation function may
depend on ``$t$'' as
\begin{eqnarray}
  \label{eq:noise-correlation-function-time-domain-t-tau}
  C_{AB}(t,\tau)
  =
  \frac{1}{2} \langle
  \hat{A}(t+\tau)\hat{B}(t)
  +
  \hat{B}(t)\hat{A}(t+\tau)
  \rangle
  .
\end{eqnarray}
From Eq.~(\ref{eq:noise-correlation-function-time-domain-t-tau}), we
estimate the correlation function for the stationary noise by
\begin{eqnarray}
  C_{({\rm av})AB}(\tau)
  :=
  \lim_{T\rightarrow\infty} \frac{1}{T}
  \int_{-T/2}^{T/2} dt C_{AB}(t,\tau)
  .
  \label{eq:noise-correlation-function-time-domain-t-tau-time-average}
\end{eqnarray}
We use
Eq.~(\ref{eq:noise-correlation-function-time-domain-t-tau-time-average})
of the correlation function for the stationary noise, instead of
Eq.~(\ref{eq:Kimble-noise-correlation-function-time-domain-tau}).
When $\hat{A}(t)=\hat{B}(t)$, the auto-correlation function
$C_{(av)A}(\tau)$ for ``stationary noise'' is given by
\begin{eqnarray}
  \label{eq:auto-correlation-function-time-domain-t-tau-time-average}
  C_{({\rm av})A}(\tau)
  :=
  \lim_{T\rightarrow\infty} \frac{1}{T} \int_{-T/2}^{T/2} dt
  C_{AA}(t,\tau)
  .
\end{eqnarray}


When the operator $A(t)$ has a non-trivial expectation value
$\langle\hat{A}(t)\rangle$, we consider the noise operator
$\hat{A}_{n}(t)$ defined by
\begin{eqnarray}
  \label{eq:Anoise-def}
  \hat{A}(t)=:\hat{A}_{n}(t)+\langle\hat{A}(t)\rangle
\end{eqnarray}
and evaluate the noise correlation function by
\begin{eqnarray}
  \label{eq:auto-correlation-time-domain-tau-time-average-SA-A-nonise}
  C_{({\rm av})A_{n}}(\tau)
  =:
  C_{({\rm av})A}(\tau)
  -
  C_{({\rm av},{\rm cl})A}(\tau)
  ,
\end{eqnarray}
where $C_{({\rm av},{\rm cl})A}(\tau)$ is the classical correlation
defined by
\begin{eqnarray}
  \label{eq:classical-correlation-function-def}
  C_{({\rm av},{\rm cl})A}(\tau)
  :=
  \lim_{T\rightarrow\infty} \frac{1}{T} \int_{-T/2}^{T/2} dt
  \langle\hat{A}(t+\tau)\rangle\langle\hat{A}(t)\rangle
  .
\end{eqnarray}
The noise spectral density $S_{A_{n}}(\omega)$ is given by the Fourier
transformation of $C_{({\rm av})A_{n}}(\tau)$ as
\begin{eqnarray}
  \label{eq:noise-spectral-density-An-def}
  S_{A_{n}}(\omega)
  :=
  \int_{-\infty}^{+\infty} d\tau
  C_{({\rm av})A_{n}}(\tau)
  e^{+i\omega\tau}
  .
\end{eqnarray}
In this Letter, we evaluate the quantum noise through the noise
spectral density (\ref{eq:noise-spectral-density-An-def}) instead of Eq.~(\ref{eq:Kimble-noise-spectral-density-double-side}).


\section{Estimation of Quantum noise}
\label{sec:Estimation_of_Quantum_noise}


In this section, we evaluate quantum noise in the case that Glauber's
number operator (\ref{eq:multi-mode-photon-number-def}) is the
directly measured operator
(Sec.~\ref{sec:Quantum_Noise_in_Balanced_Homodyne_Detections_By_PhotonNumberDetectors})
and that in the case the power operator
(\ref{eq:multi-mode-power-def}) is the directly measured operator (Sec.~\ref{sec:Quantum_Noise_in_Balanced_Homodyne_Detections_By_PowerCountingDetectors}).


In the noise estimation, we carefully treat the two types of vacuum
fluctuations from the signal field $\hat{E}_{b}$ and the field
$\hat{E}_{l_{i}}$ from the local oscillator.
From Eqs.~(\ref{eq:usual-commutation-relation-0}) and
(\ref{eq:usual-commutation-relation-vacuum}), the commutation
relations of the electric field $\hat{E}_{a}^{(\pm)}(t)$ are given by
\begin{eqnarray}
  \label{eq:commutation-relation-E+E+and-E-E-}
  \left[
  \hat{E}_{a}^{(+)}(t),\hat{E}_{a}^{(+)}(t')
  \right]
  &=&
      \left[
      \hat{E}_{a}^{(-)}(t),\hat{E}_{a}^{(-)}(t')
      \right] = 0
      ,
  \\
  \left[
  \hat{E}_{a}^{(+)}(t), \hat{E}_{a}^{(-)}(t')
  \right]
  &=&
      \frac{2\pi\hbar}{{\cal A}c}
      \int_{0}^{+\infty} \frac{d\omega}{2\pi} \omega
      e^{-i\omega(t-t')}
      \nonumber
  \\
  &=:&
       \frac{2\pi\hbar}{{\cal A}c} \Delta_{a}(t-t')
       .
       \label{eq:commutation-relation-E+E-}
\end{eqnarray}
The subscription ``$a$'' of the function $\Delta_{a}(t-t')$ indicates
the vacuum fluctuations from the electric field $\hat{E}_{a}$ with the
quadrature $\hat{a}(\omega)$.
Although the function $\Delta_{a}(t-t')$ formally diverges, we regard
that the integration range over $\omega$ in
Eq.~(\ref{eq:commutation-relation-E+E-}) as
$[\omega_{\min},\omega_{\max}]$ instead of $[0,+\infty]$.
In the actual measurement of a time sequence of a variable, we
have the minimal time bin which gives the maximal frequency
$\omega_{\max}$ and the finite whole observation time which gives the
minimum frequency $\omega_{min}$.


\subsection{Glauber's Photon-Number Detectors Case}
\label{sec:Quantum_Noise_in_Balanced_Homodyne_Detections_By_PhotonNumberDetectors}


Here, we evaluate the noise spectral density $S_{s_{Nn}}(\omega)$ for
the noise operator $\hat{s}_{Nn}(t)$ defined by
\begin{eqnarray}
  \label{eq:sNt-noise-operator}
  \hat{s}_{Nn}(t)
  :=
  \hat{s}_{N}(t)
  -
  \langle\hat{s}_{N}(t)\rangle
  .
\end{eqnarray}
The evaluation is carried out step by step.
First, we evaluate the normal-ordered noise-spectral density
$S_{S_{Nn}}^{({\rm normal})}(\omega)$, in which all vacuum
fluctuations are neglected.
Second, we evaluate the contribution from the vacuum fluctuations of
the signal field $\hat{E}_{b}(t)$.
Finally, we include the contribution from the vacuum fluctuations
from the local oscillator $\hat{E}_{l_{i}}(t)$.


\subsubsection{Normal ordered noise spectral density}
\label{sec:Normal_ordered_noise_spectral_density}


Here, we consider the normal-ordered noise spectral density
$S_{s_{Nn}}^{({\rm normal})}(\omega)$ through the normal-ordered
correlation function
\begin{eqnarray}
  \!\!\!\!\!
  C_{({\rm av})s_{N}}^{({\rm normal})}(\tau)
  \!\!\!\!&:=&\!\!\!\!
       \lim_{T\rightarrow\infty} \frac{1}{T} \int_{-T/2}^{T/2} dt
       \nonumber\\
  && \!\!\!\!\!\!
     \times
     \frac{1}{2}
     \left\langle
     :
     \hat{s}_{N}(t+\tau)\hat{s}_{N}(t)
     +
     \hat{s}_{N}(t)\hat{s}_{N}(t+\tau)
     :
     \right\rangle
     .
     \label{eq:normal-ordered-correlation-function-number-def}
\end{eqnarray}
Through Eqs.~(\ref{eq:hatsNt-electric-field}),
(\ref{eq:sNt-noise-operator}),
(\ref{eq:normal-ordered-correlation-function-number-def}), its noise
spectral density is
\begin{eqnarray}
  S_{s_{Nn}}^{({\rm normal})}(\omega)
  \!\!\!\!&=&\!\!\!\!
      \frac{{\cal A}c}{2\pi\hbar} \omega_{0} |\gamma|^{2}
      \int_{-\infty}^{+\infty} d\tau e^{+i\omega\tau}
      \lim_{T\rightarrow\infty} \frac{1}{T} \int_{-T/2}^{T/2} dt
      \nonumber\\
  &&
     \times
     \left[
     e^{-2i\theta} e^{+i\omega_{0}(2t+\tau)}
     \left\langle
     \hat{E}_{bn}^{(+)}(t+\tau) \hat{E}_{bn}^{(+)}(t)
     \right\rangle
      \right.
      \nonumber\\
  && \quad
     \left.
      +
      e^{-i\omega_{0}\tau}
      \left\langle
      \hat{E}_{bn}^{(-)}(t+\tau) \hat{E}_{bn}^{(+)}(t)
      \right\rangle
      \right.
      \nonumber\\
  && \quad
     \left.
      +
      e^{+i\omega_{0}\tau}
      \left\langle
      \hat{E}_{bn}^{(-)}(t) \hat{E}_{bn}^{(+)}(t+\tau)
      \right\rangle
      \right.
      \nonumber\\
  && \quad
     \left.
      +
      e^{-i\omega_{0}(2t+\tau)}
      \left\langle
      \hat{E}_{bn}^{(-)}(t+\tau) \hat{E}_{bn}^{(-)}(t)
      \right\rangle
      \right]
      .
     \label{eq:normal-ordered-noise-spectral-density-monochro}
\end{eqnarray}


Here, we introduce the Fourier transformed expression of the field
operator $\hat{E}_{bn}^{(\pm)}(t)$ with the noise quadrature
$\hat{b}_{n}(\omega):=\hat{b}(\omega)-\langle\hat{b}(\omega)\rangle$
as
Eq.~(\ref{eq:electric-field-notation-total-electric-field-positive}).
Furthermore, to evaluate
Eq.~(\ref{eq:normal-ordered-noise-spectral-density-monochro}),
we use the fact that the measure of the function
\begin{eqnarray}
  f(a)
  \!\!\!\!&:=&\!\!\!\!
       \lim_{T\rightarrow+\infty} \frac{1}{T} \int_{-T/2}^{T/2}dt
       e^{-iat}
       =
       \left\{
       \begin{array}{lcl}
         0 & \mbox{for} & a\neq 0, \\
         1 & \mbox{for} & a= 0.
       \end{array}
       \right.
\end{eqnarray}
in the integration over $a$ is zero, except for the case where the
delta-function $\delta(a)$ is included in the
integration~\cite{K.Nakamura-BHD-Multi-Full-Paper-in-preparation}.


If we consider the Michelson interferometer as an explicit example of
the main interferometer, each term in
Eq.~(\ref{eq:normal-ordered-noise-spectral-density-monochro}) gives
the finite value due to the appearance of the delta-function in the
expectation value of the product of the noise quadrature $\hat{b}_{n}$
and the average-integral by $t$.
These finite results of each term in
Eq.~(\ref{eq:normal-ordered-noise-spectral-density-monochro}) also
implies that if we omit the integration by the frequency and specify
the frequency so that the exponent in the averaged function vanishes,
by hand, the factor of the delta-function appears in
Eq.~(\ref{eq:normal-ordered-noise-spectral-density-monochro}).
This is the imposition of the stationarity to the noise spectral density.
In other words, for stationary noise, the correct result is obtained
if we regard the expression of
Eq.~(\ref{eq:normal-ordered-noise-spectral-density-monochro}) as
\begin{eqnarray}
  &&
     \!\!\!\!\!\!\!\!\!\!\!\!\!\!\!\!\!\!\!\!\!
     2 \pi \delta(\omega-\omega')
     S_{s_{Nn}}^{({\rm normal})}(\omega)
     \nonumber\\
  &\sim&
         \omega_{0}^{2} |\gamma|^{2}
         \left\langle
         e^{-2i\theta} \hat{b}_{n}(\omega_{0}+\omega) \hat{b}_{n}(\omega_{0}-\omega')
         \right.
         \nonumber\\
  && \quad\quad\quad\quad
     \left.
     +
     \hat{b}_{n}^{\dagger}(\omega_{0}-\omega) \hat{b}_{n}(\omega_{0}-\omega')
     \right.
     \nonumber\\
  && \quad\quad\quad\quad
     \left.
     +
     \hat{b}_{n}^{\dagger}(\omega_{0}+\omega) \hat{b}_{n}(\omega_{0}+\omega')
     \right.
     \nonumber\\
  && \quad\quad\quad\quad
     \left.
     +
     e^{+2i\theta} \hat{b}_{n}^{\dagger}(\omega_{0}+\omega) \hat{b}_{n}^{\dagger}(\omega_{0}-\omega')
     \right\rangle
     .
     \label{eq:2pideltaSNsnormal-expression}
\end{eqnarray}
Here, we used the situation $\omega_{0}\gg\omega>0$.
Note that $\omega_{0}$ is the central frequency of local oscillator
and may not coincide with the central frequency of the signal field
$\hat{E}_{b}(t)$.


Here, we regard that the central frequency $\omega_{0}$ of the local
oscillator coincides with the central frequency from the main interferometer.
This is the ``{\it homodyne detection}.''
Then, we may use the sideband picture
$\hat{b}_{\pm}(\omega):=\hat{b}(\omega_{0}\pm\omega)$ and introduce
the noise quadratures $\hat{b}_{1n}$, $\hat{b}_{2n}$, and
$\hat{b}_{\theta}$ as Eqs.~(\ref{eq:DBHD_20180805_1.1}) and
(\ref{eq:hatb1-hatb2-def})
\begin{eqnarray}
  &&
     \!\!\!\!\!\!\!\!\!\!\!\!\!\!\!\!\!\!\!\!\!
     2 \pi \delta(\omega-\omega')
     S_{s_{Nn}}^{({\rm normal})}(\omega)
     \nonumber\\
  &\sim&
         \omega_{0}^{2} |\gamma|^{2}
         \left[
         \left\langle
         \hat{b}_{\theta n}^{\dagger}(\omega')
         \hat{b}_{\theta n}(\omega)
         +
         \hat{b}_{\theta n}(\omega)
         \hat{b}_{\theta n}^{\dagger}(\omega')
         \right\rangle
         \right.
         \nonumber\\
  && \quad\quad\quad\quad
     \left.
     -
     2 \pi \delta(\omega-\omega')
     \right]
     .
     \label{eq:2pideltaSNsnormal-expression-homodyne-btheta}
\end{eqnarray}
Here, we note that
$\left[\hat{b}_{\theta n}(\omega),\hat{b}_{\theta n}(\omega')\right]=0$.
Since we only consider the positive frequency $\omega$ by the
definition
(\ref{eq:electric-field-notation-total-electric-field-positive}), the
first term in the right-hand side of
Eq.~(\ref{eq:2pideltaSNsnormal-expression-homodyne-btheta}) coincides
with Kimble's single-sideband noise spectral density
(\ref{eq:Kimble-noise-spectral-density-single-side}):
\begin{eqnarray}
  \label{eq:2pideltaSNsnormal-expression-homodyne-btheta-vs-Kimble}
  S_{s_{Nn}}^{({\rm normal})}(\omega)
  &\sim&
         \omega_{0}^{2} |\gamma|^{2}
         \left[
         \bar{S}_{b_{\theta}}^{(s)}(\omega)
         -
         1
         \right]
         .
\end{eqnarray}


\subsubsection{Vacuum fluctuations from the main interferometer}
\label{sec:Vacuum_fluctuations_from_main}


Here, we clarify the contribution of the vacuum fluctuations from the
signal field $\hat{E}_{b}(t)$ through ignoring the vacuum fluctuations
of the local oscillator $\hat{E}_{l_{i}}(t)$.
To carry out this, we evaluate
$\langle\hat{s}_{Nn}(t)\hat{s}_{Nn}(t+\tau)+\hat{s}_{Nn}(t+\tau)\hat{s}_{Nn}(t)\rangle$
under the premises
\begin{eqnarray}
  &&
     \label{eq:commutation-relation-E+E-b-nonvanish}
     \left[
     \hat{E}_{b}^{(+)}(t), \hat{E}_{b}^{(-)}(t')
     \right]
     =:
     \frac{2\pi\hbar}{{\cal A}c} \Delta_{b}(t-t')
     \neq
     0
     ,
  \\
  &&
     \label{eq:commutation-relation-E+E-li-vanish}
     \left[
     \hat{E}_{l_{i}}^{(+)}(t), \hat{E}_{l_{i}}^{(-)}(t')
     \right]
     =:
     \frac{2\pi\hbar}{{\cal A}c} \Delta_{l_{i}}(t-t')
     =
     0
     .
\end{eqnarray}
Eq.~(\ref{eq:commutation-relation-E+E-li-vanish}) is not
consistent with quantum field theory, but we dare to use these
premises (\ref{eq:commutation-relation-E+E-b-nonvanish}) and
(\ref{eq:commutation-relation-E+E-li-vanish}) to distinguish the
contribution of vacuum fluctuations from $\hat{E}_{b}$ and
$\hat{E}_{l_{i}}$.
Then, the time-averaged correlation function is given by
\begin{eqnarray}
  \label{eq:sNn-correlation-normal-sig.vac.-is-normal+sig.vac.}
  C_{({\rm av})s_{Nn}}^{({\rm normal}+{\rm sig.vac.})}(\tau)
  =
  C_{({\rm av})s_{N}}^{({\rm normal})}(\tau)
  +
  C_{({\rm av})s_{N}}^{({\rm sig.vac.})}(\tau)
  ,
\end{eqnarray}
where
\begin{eqnarray}
  C_{({\rm av})s_{N}}^{({\rm sig.vac.})}(\tau)
  \!\!\!\!&:=&\!\!\!\!
       \lim_{T\rightarrow+\infty} \frac{1}{T}
       \int_{-T/2}^{T/2} dt \frac{1}{2} \left(
       \gamma^{*}(t+\tau) \gamma(t) \Delta_{b}(\tau)
       \right.
       \nonumber\\
  && \quad\quad\quad\quad\quad
     \left.
     +
     \gamma^{*}(t) \gamma(t+\tau) \Delta_{b}(-\tau)
     \right)
     .
     \label{eq:sNn-correlation-sig.vac.-def}
\end{eqnarray}
In the monochromatic local oscillator case
(\ref{eq:monochromatic-gamma-omega}),
we obtain
\begin{eqnarray}
  \label{eq:sNn-correlation-sig.vac.-monochro-local}
  C_{({\rm av})s_{Nn}}^{({\rm sig.vac.})}(\tau)
  \!\!\!\!&=&\!\!\!\!
      \frac{1}{2} \omega_{0} |\gamma|^{2} \left(
      e^{+i\omega_{0}\tau} \Delta_{b}(\tau)
      +
      e^{-i\omega_{0}\tau} \Delta_{b}(-\tau)
      \right)
      ,
  \\
  \label{eq:sNn-noise-spectral-sig.vac.-monochro-local}
  S_{s_{Nn}}^{({\rm sig.vac.})}(\omega)
  \!\!\!\!&=&\!\!\!\!
      \omega_{0}^{2} |\gamma|^{2}
  ,
\end{eqnarray}
in the situation $\omega_{0}\gg\omega>0$.
Together with
Eq.~(\ref{eq:2pideltaSNsnormal-expression-homodyne-btheta-vs-Kimble}),
we obtain
\begin{eqnarray}
  \label{eq:2pideltaSNsnormal+sig.vac.-expression-homodyne-btheta-vs-Kimble}
  S_{s_{Nn}}^{({\rm normal}+{\rm sig.vac.})}(\omega)
  =
  \omega_{0}^{2} |\gamma|^{2}
  \bar{S}_{b_{\theta}}^{(s)}(\omega)
  .
\end{eqnarray}
Thus, apart from the overall factor, this result coincides with
Kimble's single-sideband noise spectral density
(\ref{eq:Kimble-noise-spectral-density-single-side}).
This means that the Kimble's single-sideband noise spectral density is
already included the contribution of the vacuum fluctuations from the
signal field $\hat{E}_{b}$.


\subsubsection{Vacuum fluctuations from the local oscillator}
\label{sec:Vacuum_fluctuations_from_local_oscillator}


Here, we consider the vacuum fluctuations from the local oscillator
$\hat{E}_{l_{i}}(t)$ through the premise
\begin{eqnarray}
  \label{eq:Deltalineq0-Deltabneq0}
  \left[
  \hat{E}_{l_{i}}^{(+)}(t), \hat{E}_{l_{i}}^{(-)}(t')
  \right]
  =
  \frac{2\pi\hbar}{{\cal A}c} \Delta_{l_{i}}(t-t')
  \neq
  0
\end{eqnarray}
instead of the premise (\ref{eq:commutation-relation-E+E-li-vanish}).
Then, we obtain
\begin{eqnarray}
  \label{eq:total-averaged-correlation-function}
  C_{({\rm av})s_{Nn}}(\tau)
  =
  C_{({\rm av})s_{Nn}}^{({\rm normal}+{\rm sig.vac.})}(\tau)
  +
  C_{({\rm av})s_{Nn}}^{({\rm loc.vac.})}(\tau)
  ,
\end{eqnarray}
and
\begin{eqnarray}
  &&
     \!\!\!\!\!\!\!\!\!\!\!\!\!\!\!\!\!
     S_{s_{Nn}}^{({\rm loc.vac.})}(\omega)
     := \int_{-\infty}^{+\infty} d\tau e^{+i\omega\tau} C_{({\rm av})s_{Nn}}^{({\rm loc.vac.})}(\tau)
     \nonumber\\
  &=&
      \frac{1}{2}
      \int_{0}^{+\infty} \frac{d\omega_{1}}{2\pi}
      \int_{0}^{+\infty} \frac{d\omega_{2}}{2\pi}
      \sqrt{\omega_{1}\omega_{2}}
      \left\langle
      \hat{b}_{n}^{\dagger}(\omega_{1})
      \hat{b}_{n}(\omega_{2})
      \right\rangle
      \nonumber\\
  &&
     \quad
     \times
      \left[
      \Theta(\omega_{2}-\omega) (\omega_{2}-\omega)
      +
      \Theta(\omega_{2}+\omega) (\omega_{1}+\omega)
      \right]
      \nonumber\\
  &&
     \quad
     \times
     \lim_{T\rightarrow+\infty} \frac{1}{T}
     \int_{-T/2}^{T/2}dt e^{+i(\omega_{1}-\omega_{2})t}
     .
     \label{eq:averaged-noise-spectral-density-from-loc.vac.}
\end{eqnarray}
Here, we used the mode expansion expression of
$\hat{E}^{(\pm)}_{b_{n}}(t)$ and ``{\it complete dark port
  condition}'' which means the absence of the delta-function
$\delta(\omega-\omega_{0})$ in the expectation value
$\left\langle\hat{b}(\omega)\right\rangle$.


Here, we introduce a new noise-spectral density
$\SF_{b_{n}}(\omega)$ by
\begin{eqnarray}
  &&
     \!\!\!\!\!\!\!\!\!\!\!\!\!\!
     2 \pi \delta(\omega_{1}-\omega_{2})
     \SF_{b_{n}}(\omega_{1})
     :=
     \left\langle
     \hat{b}_{n}(\omega_{1}) \hat{b}_{n}^{\dagger}(\omega_{2})
     +
     \hat{b}_{n}^{\dagger}(\omega_{2}) \hat{b}_{n}(\omega_{1})
     \right\rangle
     .
     \nonumber\\
  \label{eq:calS-bn-omega-def}
\end{eqnarray}
This definition of $\SF_{b_{n}}(\omega)$ has the same form of
the Kimble single-sideband noise spectral density
(\ref{eq:Kimble-noise-spectral-density-single-side}).
However, the noise-spectral density $\SF_{b_{n}}(\omega)$ have nothing
to do with the two-photon formulation.
The frequencies $\omega_{1}$ and $\omega_{2}$ in
Eq.~(\ref{eq:calS-bn-omega-def}) is not sideband frequencies, but the
frequency $\omega$ in Eq.~(\ref{eq:electric-field-notation-total-electric-field-positive}).
Through Eq.~(\ref{eq:calS-bn-omega-def}) and $\omega>0$, we obtain
\begin{eqnarray}
  S_{s_{Nn}}(\omega)
  \!\!\!\!&:=&\!\!\!\!
               S_{s_{Nn}}^{({\rm normal}+{\rm sig.vac.})}(\omega)
               +
               S_{s_{Nn}}^{({\rm loc.vac.})}(\omega)
               \nonumber\\
          &\sim&
                 \omega_{0}^{2} |\gamma|^{2}
                 \bar{S}_{b_{\theta}}^{(s)}(\omega)
                 \nonumber\\
          &&
             +
             \frac{1}{2}
             \int_{0}^{+\infty} \frac{d\omega_{1}}{2\pi}
             (\omega_{1})^{2}
             \left(
             \SF_{b_{n}}(\omega_{1}) - 1
             \right)
             \nonumber\\
          &&
             -
             \frac{1}{4}
             \int_{0}^{\omega} \frac{d\omega_{1}}{2\pi}
             \omega_{1}(\omega_{1}-\omega)
             \left(
             \SF_{b_{n}}(\omega_{1}) - 1
             \right)
             .
             \label{eq:sNn-noise-spectral-density-total-full}
\end{eqnarray}
Here, we note that Kimble's noise spectral density is realized when
$|\gamma|^{2}$ is sufficiently large so that the second- and
third lines in the right-hand side of
Eq.~(\ref{eq:sNn-noise-spectral-density-total-full}) are negligible,
while the effects of these lines appears when $|\gamma|^{2}$ is
sufficiently small.
These lines are contribution of the vacuum fluctuations from the local
oscillator and the third line has the frequency dependence.


\subsection{Power Counting Detectors Case}
\label{sec:Quantum_Noise_in_Balanced_Homodyne_Detections_By_PowerCountingDetectors}


To evaluate the noise spectral density $S_{s_{Pn}}(\omega)$ of the
operator
\begin{eqnarray}
  \label{eq:sPn-def}
  \hat{s}_{Pn}:=\hat{s}_{P}-\langle\hat{s}_{P}\rangle,
\end{eqnarray}
we consider the noise correlation function
\begin{eqnarray}
  C_{s_{P_{n}}}(t,\tau)
  \!\!\!\!&:=&\!\!\!\!
               \frac{1}{2}
               \left\langle
               \hat{s}_{P_{n}}(t+\tau)\hat{s}_{P_{n}}(t)
               +
               \hat{s}_{P_{n}}(t)\hat{s}_{P_{n}}(t+\tau)
               \right\rangle
               ,
               \label{eq:noise-correlation-function-sPn-def}
  \\
  \!\!\!\!&=&\!\!\!\!
              \frac{1}{2}
              \left(
              \gamma(t+\tau) + \gamma^{*}(t+\tau)
              \right)
              \left(
              \gamma(t) + \gamma^{*}(t)
              \right)
              \nonumber\\
          && \quad
             \times
             \left\langle
             \sqrt{\frac{{\cal A}c}{2\pi\hbar}} \hat{E}_{bn}(t+\tau)
             \sqrt{\frac{{\cal A}c}{2\pi\hbar}} \hat{E}_{bn}(t)
             \right.
             \nonumber\\
          && \quad\quad\quad
             \left.
             +
             \sqrt{\frac{{\cal A}c}{2\pi\hbar}} \hat{E}_{bn}(t)
             \sqrt{\frac{{\cal A}c}{2\pi\hbar}} \hat{E}_{bn}(t+\tau)
             \right\rangle
             \nonumber\\
          &&\!\!\!\!
             +
             \frac{1}{2}
             \Delta_{l_{i}}(-\tau)
             \left\langle
             \sqrt{\frac{{\cal A}c}{2\pi\hbar}} \hat{E}_{b}(t)
             \sqrt{\frac{{\cal A}c}{2\pi\hbar}} \hat{E}_{b}(t+\tau)
             \right\rangle
             \nonumber\\
          &&\!\!\!\!
             +
             \frac{1}{2}
             \Delta_{l_{i}}(t)
             \left\langle
             \sqrt{\frac{{\cal A}c}{2\pi\hbar}} \hat{E}_{b}(t+\tau)
             \sqrt{\frac{{\cal A}c}{2\pi\hbar}} \hat{E}_{b}(t)
             \right\rangle
             .
             \label{eq:CsPttau-explicit}
\end{eqnarray}
The averaged noise correlation function $C_{({\rm av})s_{Pn}}$ and the
noise spectral density $S_{s_{Pn}}(\omega)$ are given by the similar
manner to Glauber's photon-number case
(\ref{eq:sNn-noise-spectral-density-total-full}) through
Eqs.~(\ref{eq:auto-correlation-function-time-domain-t-tau-time-average})
and (\ref{eq:noise-spectral-density-An-def}).
Then, we have
\begin{eqnarray}
  S_{s_{Pn}}(\omega)
  \!\!\!\!&=&\!\!\!\!
      \omega_{0}^{2} |\gamma|^{2}
      \bar{S}_{b_{\theta}}^{(s)}(\omega)
      \nonumber\\
  &&
     +
     \frac{1}{2} \int_{0}^{\infty} \frac{d\omega_{1}}{2\pi}
     \omega_{1}^{2}
     \left(
     \SF_{b_{n}}(\omega_{1})
     -
     1
     \right)
      \nonumber\\
  &&
     +
     \frac{1}{2}
     \int_{0}^{\omega} \frac{d\omega_{1}}{2\pi}
     \omega_{1}
     (\omega-\omega_{1})
     \SF_{b_{n}}(\omega_{1})
     ,
     \label{eq:noise-spectral-power-counting-general-result}
\end{eqnarray}
in the situation $\omega_{0}\gg\omega>0$.
We also note that even in the noise spectral density
(\ref{eq:noise-spectral-power-counting-general-result}) Kimble's noise
spectral density is realized when $|\gamma|^{2}$ is sufficiently large
so that the second- and third lines in the right-hand side of
Eq.~(\ref{eq:noise-spectral-power-counting-general-result}) is
negligible, while these effects appears when $|\gamma|^{2}$ is
sufficiently small.
These lines are contribution from the vacuum fluctuations from the
local oscillator.
We note that the third line in
Eq.~(\ref{eq:noise-spectral-power-counting-general-result}), which is
the frequency dependent term, is different from that in
Eq.~(\ref{eq:sNn-noise-spectral-density-total-full}).


\section{Summary}
\label{sec:Summary}


In summary, we showed our estimation of quantum noise in the balanced
homodyne detections which measure $\hat{b}_{\theta}(\omega)$ as the
expectation value.
We consider both cases in which the directly measured operators at the
photodetectors are Glauber's photon-number operator
(\ref{eq:multi-mode-photon-number-def}) and the power operator
(\ref{eq:multi-mode-power-def}), respectively.
In our estimation, we did not use the two-photon formulation which is
widely used in the gravitational-wave community.
We concentrate on the stationary noise of the system through the
time-average procedure.
We also carefully treat vacuum fluctuations in our noise estimation.
In both cases, we have derived the deviations from the Kimble's noise
spectral density (\ref{eq:Kimble-noise-spectral-density-single-side}),
which are beyond the two-photon formulation.


In general, such deviations from the Kimble's noise spectral density
may also occur due to the physical properties of the photodetectors
such as the band structure of photodiodes.
However, the noise spectral densities derived here are for ideal
Glauber's photon-number counting detectors or ideal power counting
detectors.
The obtained deviations are due to the vacuum fluctuations from the
local oscillator.
The derived noise spectral densities
(\ref{eq:sNn-noise-spectral-density-total-full}) and
(\ref{eq:noise-spectral-power-counting-general-result}) yield the
Kimble noise spectral density when the amplitude $|\gamma|$ of the
coherent state from the local oscillator is sufficiently large.
On the other hand, when the amplitude of the coherent state from the
local oscillator is sufficiently small, the difference of these noise
spectral appears.
This will be able to use for the characterization of the ideal
multi-mode photon number detectors or ideal multi-mode photon power
detectors.


In the case where the directly measured operator of the photodetector
is the number operator of each frequency modes in
Refs.~\cite{K.Nakamura-M.-K.Fujimoto-double-balanced},
we reached to the conclusion that the measurement of the expectation
value of the operator $\hat{b}_{\theta}(\omega)$ by the balanced
homodyne detection is impossible.
Therefore, we had to consider the eight-port homodyne detection which
enable us to measure the expectation value of the operator
$\hat{b}_{\theta}(\omega)$.
Together with the ingredients of this Letter, this indicates that the
choice of the directly measured operator at the photodetector affects
the result not only of the noise properties but also the output signal
expectation values themselves.
Therefore, we conclude that the specification of the directly
measured operator is crucial in the development of quantum
measurement theory.
Details of the derivation of our formulae will be seen in
elsewhere~\cite{K.Nakamura-BHD-Multi-Full-Paper-in-preparation}.




\section*{Acknowledgments}


The author deeply acknowledges to Prof. Masa-Katsu Fujimoto for
valuable comments and his continuous encouragements.
The author also acknowledges to Prof. Takayuki Tomaru, and
Prof. Tomotada Akutsu, Prof. Shinji Miyoki (ICRR, Tokyo Univ.), and
Prof. Osamu Miyakawa (ICRR, Tokyo Univ.), and the other members of the
GWSP in NAOJ for their continuous encouragement and discussions to
this research.


\section*{References}

\end{document}